\documentclass[prd,aps,twocolumn,amsfonts,showpacs,nofootinbib,preprintnumbers,superscriptaddress]{revtex4-1}

\usepackage{amsmath,amsthm,amssymb} 
\usepackage{dsfont,yfonts}
\usepackage{hyperref}
\usepackage{floatrow,dcolumn,rotating}
\usepackage{array,xcolor,graphicx}

\usepackage{graphicx}
\usepackage{tikz}
 \usetikzlibrary{decorations.pathmorphing,decorations.markings,trees,positioning,arrows}
 \usetikzlibrary{calc,patterns}
 \usetikzlibrary{shapes,backgrounds}   
 \tikzset{
    photon/.style={decorate, decoration={snake}, draw=black},
    particle/.style={draw=black,postaction={decorate,
        decoration={markings,mark=at position 0.65 with {\arrow[draw=black]{>}}}}},
    antiparticle/.style={draw=black,postaction={decorate,
        decoration={markings,mark=at position 0.5 with {\arrow[draw=black]{<}}}}},
    gluon/.style={decorate, draw=orange,very thick, 
	    decoration={coil,amplitude=4pt, segment length=6pt}},
    higgs/.style={draw=red,very thick, postaction={decorate},
	           decoration={markings,mark=at position .55 with  {\arrow[draw=red]{>}}}},
     doubleline/.style={draw=black,very thick,
            double distance=8pt,
            postaction={decorate,
            decoration={
                markings,
                mark=between positions 10pt and -10pt step 50pt with {
                    \arrow[very thick,yshift= 4pt,xshift=.8pt]{>}
                    \arrow[very thick,yshift=-4pt,xshift=.8pt]{<}
                },
            }},
        }}

\hypersetup{colorlinks=true,linkcolor=blue,citecolor=red,urlcolor=blue}

\newcommand{\nn}{\nonumber\\}

\newcommand{\cG}{{\cal G}}
\renewcommand{\vec}[1]{\mathbf{#1}}

\def\rar{\rightarrow}
\def\la{\langle}
\def\ra{\rangle}
\def\Tr{{\mathrm{Tr}}}
\def\tr{{\mathrm{tr}}}

\def\CE{\mathcal{E}}

\def\CL{\mathcal{L}}

\def\CN{\mathcal{N}}

\begin{document}
\preprint{MIT-CTP/5004}

\title{Kinetic theory for classical and quantum many-body chaos}

\author{Sa\v{s}o Grozdanov}
\affiliation{Center for Theoretical Physics, MIT, Cambridge, MA 02139, USA}
\author{Koenraad Schalm}
\affiliation{Instituut-Lorentz for Theoretical Physics $\Delta$ITP,
  Leiden University, Niels Bohrweg 2, Leiden 2333 CA, The Netherlands}
\author{Vincenzo Scopelliti}
\affiliation{Instituut-Lorentz for Theoretical Physics $\Delta$ITP, Leiden University, Niels Bohrweg 2, Leiden 2333 CA, The Netherlands}

\begin{abstract} 
\noindent
For perturbative scalar field theories, the late-time-limit of the out-of-time-ordered correlation function that measures (quantum) chaos is shown to be equal to a Boltzmann-type kinetic equation that measures the total gross (instead of net) particle exchange between phase space cells, weighted by a function of energy. This derivation gives a concrete form to numerous attempts to derive chaotic many-body dynamics from ad hoc kinetic equations.  A period of exponential growth in the total gross exchange determines the Lyapunov exponent of the chaotic system. Physically, the exponential growth is a front propagating into an unstable state in phase space. As in conventional Boltzmann transport, which follows from the dynamics of the net particle number density exchange, the kernel of this kinetic integral equation for chaos is also set by the 2-to-2 scattering rate. This provides a mathematically precise statement of the known fact that in dilute weakly coupled gases, transport and scrambling (or ergodicity) are controlled by the same physics. 
\end{abstract}

\maketitle
\begingroup
\hypersetup{linkcolor=black}
\tableofcontents
\endgroup

\section{Introduction}
The weakly interacting dilute gas is one of the pillars of physics. It provides a canonical example for the statistical foundation of thermodynamics and its kinetic description---the Boltzmann equation---allows for a computation of the collective transport properties from collisions of the microscopic constituents. Historically, this provided the breakthrough evidence in favor of the molecular theory of matter. A crucial point in Boltzmann's kinetic theory is the assumption of {\em molecular chaos} whereby all $n > 2$ quasi-particle correlations are irrelevant due to diluteness and the validity of ensemble averaging, i.e. ergodicity \cite{chapman-book,kvasnikov-book,saint-raymond-book,ferziger-kaper-book,ford-book,DeGroot-book,silin-book,grad-1963,gross-1959}. However, finding a precise quantitative probe of this underlying chaotic behavior in many-body systems has been a notoriously difficult problem. In the past, phenomenological approaches positing a Boltzmann-like kinetic equation (see e.g. \cite{1999chao.dyn..9034V}) have reproduced
numerically computed properties of chaos, such as the Lyapunov
exponents, but a fundamental
origin supporting this approach is lacking.

A measure of chaos applicable to both weakly coupled (kinetic) and
strongly coupled quantum systems (without quasi-particles) is a period of exponential growth of a thermal out-of-time-ordered correlator (OTOC):
\begin{align}\label{eq:1}
C(t) = \theta(t)\, \langle [\hat{W}(t),\hat{V}(0)]^\dagger[\hat{W}(t),\hat{V}(0)]\rangle_{\beta} \,,
\end{align}
where $W(t)$ and $V(0)$ are generic operators and $\beta=1/T$. For example, choosing 
$W(t)=q(t)$,~$V(0)=p(0)\equiv -i\hbar\frac{\partial}{\partial q(0)}$
one immediately sees that $C(t)$ probes the
dependence on initial conditions---and, hence, if this
dependence displays exponential growth, chaos. This OTOC was first put forth in
studies of quantum electron transport in weakly disordered materials
\cite{1997PhRvE..55.1243A,1996PhRvB..5414423A,1969JETP...28.1200L}, which noted that in quantum systems the regime of classical
  exponential growth cuts off at the so-called Ehrenfest time, and of late, it has been used to detect exponential growth of perturbations characteristic
of chaos in strongly coupled quantum systems
\cite{Shenker:2013pqa,Roberts:2014isa,Maldacena:2015waa,Polchinski:2015cea}.
This has led in turn to a reconsideration of this OTOC in weakly
coupled field theories
\cite{Stanford:2015owe,Aleiner:2016eni,Patel:2017vfp,Chowdhury:2017jzb,Roberts:2016wdl,Kukuljan:2017xag,Rozenbaum}. 
A strong impetus for this renewed interest has been a possible
connection between chaotic behavior and transport, in particular,
late-time diffusion (see e.g. recent
\cite{Blake:2017qgd,Rakovszky:2017qit,Grozdanov:2017ajz,Khemani:2017nda,Lucas:2017ibu}). Many
weakly coupled studies have indeed found such a
connection. Intuitively, this should not be a surprise. In weakly
coupled particle-like theories, chaotic short-time behavior is clearly
set by successive uncorrelated 2-to-2 scatterings, but the dilute
molecular chaos assumption in Boltzmann's kinetic theory shows that
2-to-2 scattering also determines the late-time diffusive transport
coefficients. A mathematically precise relation, however, between chaos and transport in dilute perturbative systems did not exist.

In this paper, we will provide this relation. We will show that
a direct analogue of the conventional Boltzmann transport equation,
but where one traces the total gross exchange between phase space
cells weighted by an energy factor, rather than net particle number
density, computes the late-time behavior of chaos in terms of 
the exponential growth of the OTOC of a bosonic system before the Ehrenfest time.
The resemblance between the OTOC computation and kinetic equations was already noted in \cite{Stanford:2015owe}, although with a different
interpretation. Our result is explicit in the physical meaning of the
kinetic equation for chaos and makes particularly clear the relation
between chaos and transport in dilute weakly coupled theories, as the
kernel in both cases is the \mbox{2-to-2} scattering cross-section,
even though transport is a relaxational process and chaos an exponentially
divergent one. This OTOC-derived gross exchange equation shares many of the salient features of the earlier postulated chaos-determining kinetic equations \cite{PhysRevLett.80.2035,1999chao.dyn..9034V}, explaining post facto why they obtained the correct result. 

\section{Boltzmann transport and chaos from a gross energy exchange kinetic equation}

To exhibit the essence of the statement that chaos-driven ergodicity follows from a gross exchange equation analogous to the Boltzmann equation, we first construct this equation from first principles and show how it captures the exponential growth of microscopic energy-weighted exchanges due to inter-particle collisions. Then, in the next section, we derive this statement from the late-time limit of the OTOC in perturbative quantum field theory.

Consider the linearized Boltzmann equation for the time dependence of the change of particle number density per unit of phase space: $\delta n(t,\vec{p})=n (t,\vec{p}) - n (E_\vec{p})$, where $n(\vec{p})$ is the equilibrium Bose-Einstein distribution $n(\vec{p}) = 1/(e^{\beta E(\vec{p})}-1)$ that depends on the energy $E(\vec{p})$.\footnote{For simplicity, we assume spatial homogeneity of the gas with the energy $E(\vec{p})$ and think of all quantities as averaged over space, e.g. $n(t,\vec{p})=\int\! d\vec{x}\, n(t,\vec{x},\vec{p})$.}
In terms of the one-particle distribution function, $f(t,\vec{p}) =
\frac{\delta n(t,\vec{p})}{n(\vec{p})(1+n(\vec{p}))}$ , the linearized Boltzmann
equation is a homogeneous evolution equation for
$f(t,\vec{p})$ (see e.g. \cite{SmithJensen,Balescu,HaugJauho}):\footnote{In relativistic theories $\int\displaylimits_{\vec p}\!\!\equiv\!\!
  \int \frac{d^3 p}{(2\pi)^3}\frac{1}{2E(\vec{p})}$ and \mbox{$\int_{p} \equiv \int
  \frac{d^4 p}{(2\pi)^4 }$}. For a non-relativistic system, $\int\displaylimits_{\vec p} \equiv
  \int \frac{d^3 p}{(2\pi)^3}$, and similarly, $\int_{\vec x} \equiv \int d^3 x$ and $\int_x \equiv \int d^4 x$. }
\begin{align}
  \label{eq:2}
  \partial_t f(t, \vec{p}) =-\int_{\vec{l}} {\cal  L}(\vec{p},\vec{l})f(t,\vec{l}) ,
\end{align}
where the kernel of the collision integral 
\begin{align}
{\cal L}(\vec{p},\vec{l})  \equiv - \left[R^\wedge(\vec{p},\vec{l})-R^\vee(\vec{p},\vec{l})\right]
\end{align} 
measures the difference between the rates of scattering into the
phase-space cell and scattering out the phase space cell. The factor
\begin{align}
  \label{eq:3gain}
 R^\wedge(\vec{p},\vec{l}) =&\,  \frac{1}{n(\vec{p})(1+n(\vec{p})) } \int\displaylimits_{\vec{p}_2,\vec{p}_3,\vec{p}_4}  d\Sigma  (\vec{p},\vec{p}_2|\vec{p}_3,\vec{p}_4) \nn
  &\times \left(\delta(\vec{p}_3-\vec{l})+\delta(\vec{p}_4-\vec{l})\right) ,
\end{align}
counts increases of the local density by one unit. The factor
\begin{align}
  \label{eq:3loss}
   R^\vee(\vec{p},\vec{l}) =&\, \frac{1}{n(\vec{p})(1+n(\vec{p}))} \int\displaylimits_{\vec{p}_2,\vec{p}_3,\vec{p}_4} d\Sigma (\vec{p},\vec{p}_2|\vec{p}_3,\vec{p}_4) \nn
  & \times \left(\delta(\vec{p}-\vec{l})+\delta(\vec{p}_2-\vec{l})\right) ,
\end{align}
counts decreases of the number density by one unit. Here, 
\begin{align}\label{sigma}
{d\Sigma}(\vec{p},\vec{p}_2 | \vec{p}_3,\vec{p}_4) =&\,
\, n(\vec{p}) \, n(\vec{p}_2) \, \frac{1}{2} |{\cal T}_{pp_2\rar p_3p_4}|^2 \nn
&\times(1+n(\vec{p}_3))(1+n(\vec{p}_4)) \nn
&\times (2\pi)^4\delta^4(p+p_2-p_3-p_4)
\end{align}
with $|{\cal T}_{pp_2\rar p_3p_4}|^2$ the transition amplitude squared.  By defining an inner product 
\begin{align}\label{eq:5}
\langle \phi|\psi\rangle = \int_{\vec{p}} n(\vec{p})(1+n(\vec{p})) \phi^{\ast}(\vec{p})\psi(\vec{p}) \,,
\end{align}
one can use the symmetries of the cross-section
$d\Sigma(\vec{p}_1,\vec{p}_2|\vec{p}_3,\vec{p}_4)=d\Sigma(\vec{p}_2,\vec{p}_1|\vec{p}_3,\vec{p}_4)=d\Sigma(\vec{p}_3,\vec{p}_4|\vec{p}_1,\vec{p}_2)=d\Sigma(\vec{p}_1,\vec{p}_2|\vec{p}_4,\vec{p}_3)$
to show that the operator ${\cal L}(\vec{p},\vec{l})$
is not only Hermitian on this inner product, but also positive
semidefinite---all its eigenvalues are real and $\xi_n \geq 0$. Hence,
the solutions to the Boltzmann equation are purely relaxational:
\begin{align}
f(\vec{p},t) = \sum_n A_n e^{-\xi_n t} \phi_n (\vec{p}),
\end{align} 
where $\sum_n$ formally stands for either a sum over discrete values or an integral over a continuum (see e.g. \cite{SmithJensen,Balescu,HaugJauho,Grozdanov:2016vgg}). Moreover, every $\xi =0$ eigenvalue is associated with a symmetry and has an associated conserved quantity---a collisional invariant.
  
Let us instead trace the total gross exchange, rather than the net flux,
by changing the sign of the outflow $R^\vee(\vec p, \vec l )$ in the
kernel of the integral $\CL (\vec p, \vec l )$. A distribution
function that follows from Eq. \eqref{eq:2} with the kernel ${\cal
  L}_{\scriptscriptstyle \text{total}}(\vec{p},\vec{l})=-
\left[R^\wedge(\vec{p},\vec{l}) + R^\vee(\vec{p},\vec{l})\right]$
counts additively the total in- and out-flow of particles from a
number density inside a unit of phase space. However, this
  over-counts because the loss rate $R^\vee(\vec{p},\vec{l})$ consists of a
drag (self-energy) term, $2\Gamma_{\vec{p}}$, caused by the thermal
environment --- the term proportional to $\delta(\vec{p}-\vec{l})$ in
\mbox{Eq. \eqref{eq:3loss}} --- in addition to a true loss rate term, $R^{\vee_{T}}
(\vec{p},\vec{l}) = R^{\vee} (\vec{p},\vec{l}) - 2 \Gamma_\vec{p}
\delta(\vec p - \vec l)$. Only $R^{\vee_{T}}$ changes the number of particles in $f(t,\vec{p})$ due to 
deviations coming from $f(t,\vec{p}\neq \vec{l})$. Accounting for this, and changing only the sign of the true outflow,
we arrive at a gross exchange equation
\begin{align}\label{eq:4b}
&\partial_t f_{\scriptscriptstyle \text{gross}}(t,\vec{p}) = \\
& \int_{\vec l} \left[ R^\wedge(\vec{p},\vec{l}) +  R^{\vee}(\vec{p},\vec{l})   -  4\Gamma_{\vec{p}}  \delta(\vec{p}-\vec{l}) \right] f_{\scriptscriptstyle \text{gross}}(t,\vec{l})\,. \nonumber
\end{align}

The central result of this paper is that tracking the time-evolution
  of this gross exchange---weighted additionally by an odd function
  $\CE(E)$ of the energy $E$ to be specified below---is a microscopic kinetic measure of chaos (or
scrambling). 
It is thus quantified by the distribution $f_{\scriptscriptstyle EX} \equiv \CE(E) f_{\scriptscriptstyle \text{gross}}$ and governed by
\begin{align}\label{eq:fEXQ}
&\partial_t f_{\scriptscriptstyle EX}(t,\vec{p}) =\\ \nonumber
&\int_{\vec l} \frac{\CE[E_{\vec p}]}{\CE[E_{\vec l}]} \left[ R^\wedge(\vec{p},\vec{l}) + R^\vee(\vec{p},\vec{l})-4\Gamma_{\vec{p}}  \delta(\vec{p}-\vec{l})  \right] f_{\scriptscriptstyle EX}(t,\vec{l})\,.
\end{align}
Specifically, Eq. \eqref{eq:fEXQ} can be derived from the late-time
behavior of the OTOC of local field operators in perturbative
relativistic scalar quantum field theories. The OTOC selects a specific functional $\CE(E)$, such that in the limit of high temperature, $\CE(E) \to 1 / E$. The distribution $f_{\scriptscriptstyle EX}$ can grow exponentially and indefinitely
because the Hermitian operator 
\begin{align}\label{eq:label}
\CL_{\scriptscriptstyle EX} (\vec{p},\vec{l}) =- \frac{E_{\vec{l}} }{E_{\vec{p}}}
(R^\wedge(\vec{p},\vec{l}) + R^\vee(\vec{p},\vec{l})-4\Gamma_{\vec{p}}\delta(\vec{p}-\vec{l}))
\end{align} 
is no longer positive semi-definite. It permits a set of negative
eigenvalues, $\xi_m < 0$, which characterize the exponential growth in
the amount of gross energy exchanged inside the system.  This
exponential evolution persists to $t\to\infty$
\cite{Kukuljan:2017xag}, so $\xi_m$ specify a subset of all Lyapunov
exponents $\lambda_L$ of the many-body system, with $\lambda_{L,m} =
-\xi_m $ by definition. Finally, since choosing a different odd $\CE(E)$ results in
a similarity transformation of the kernel, the spectrum of $f_{\text{OTOC}}$ equals the spectrum of $f_{\scriptscriptstyle EX}$.

The above construction tremendously simplifies the computation of the Lyapunov exponents for weakly
interacting dilute systems. Beyond providing a physically intuitive
picture of chaos, it reduces the calculation of Lyapunov exponents to
a calculation of $|{\cal T}_{pp_2\rar p_3p_4}|^2$, which is entirely
determined by particle scattering. For example, in a theory of
$N\times N$ Hermitian massive scalars $\Phi_{ab}$,
\begin{align}
\label{eq:theory}
\CL = \tr \left( \frac{1}{2}\left(\partial_t \Phi\right)^2 - \frac{1}{2} (\nabla \Phi)^2 - \frac{m^2}{2} \Phi^2 -  \frac{1}{4!} g^2 \Phi^4 \right),
\end{align}
for which the transition probability appropriately traced over external states equals 
\begin{align}
|{\cal T}_{12\rar 34}|^2\!=\! \frac{1}{6}g^4 \left (N^2 +5\right).
\end{align}
Eq. \eqref{eq:fEXQ} directly computes the Lyapunov exponents (see Fig. \ref{fig:Spect}). In the $\beta \to 0$ limit, the leading exponent becomes
\begin{align}\label{eq:7}
\lambda_{L} \simeq \frac{0.025 T^2}{48 m} \,\frac{1}{2} |{\cal T}_{12\rar 34}|^2   \simeq \frac{0.025}{4} \,  \frac{g^4(N^2+5) T^2}{144m}  \,.
\end{align}
In the large $N$ limit, Eq. \eqref{eq:7} recovers the explicit OTOC result of \cite{Stanford:2015owe} after correcting a factor of a $1/4$ miscount (see Appendix \ref{App}).

\section{A derivation of the gross exchange kinetic equation from the OTOC}

To set the stage, we first show how the linearized Boltzmann equation \eqref{eq:2} arises in quantum field theory, using the theory in Eq. \eqref{eq:theory} as an example. The derivation is closely related to the Kadanoff-Baym quantum kinetic equations \cite{DeGroot-book,altland2010condensed}. It builds on similar derivations
in \cite{Jeon:1994if,Mueller:2002gd,Jeon-Yaffe}. A
complementary approach to the derivation here, which is closer in spirit to the Kadanoff--Baym derivation, but makes the physics less transparent, is the generalized OTOC contour quantum kinetic equation of \cite{Aleiner:2016eni}. 

The one-particle distribution function $f(t, \vec{x}, \vec{p})$ follows from the Wigner
transform of the bilocal operator
\begin{align}
\label{eq:10a}
\rho(x,p)&=\int_y e^{-ip\cdot y}
\,\mathrm{Tr}\left[\Phi(x+y/2)\Phi(x-y/2)\right] \nn
&= \int_{k}
e^{ik x} \mathrm{Tr}\left[\Phi(p+k/2)\Phi(p-k/2)\right] \,.
\end{align}
When the momentum is taken to be on shell, the Wigner function
$\rho(x,p)$ becomes proportional to the relativistic one-particle
operator-valued distribution function $\rho(x,\mathbf{p},E_\mathbf{p})=
n(x,\vec p)$ \cite{DeGroot-book}. The expectation value of the scalar density is then $\la \rho \ra_\beta$.

We now consider the linearized Boltzmann equation as a dynamical
equation for fluctuations \mbox{$\delta \rho(x,p)=n(\vec{p})(1+n(\vec{p}))f(x,p)$} in the bilocal density operator:
\begin{align} \label{eq:9}
 \left[ \partial_{x^0}  \delta(x-y) \delta(p-q) + {\cal L}
  (x,p|y,q)\right] \delta\rho(y,q) = 0 \,.
\end{align}
If the fluctuations are small, and the assumption of molecular chaos
holds, the central limit theorem implies that the two point function
of the fluctuations in the bilocal density is the Green's function for the linearized Boltzmann operator 
\begin{align}
  \label{eq:10}
  &i G^{\rho\rho}_R(x,p|y,q) = \theta(x^0-y^0)\langle[\delta\rho(x,p),\delta\rho(y,q)]\rangle \nn
  &= \left[ \partial_{x^0}  \delta(x-y) \delta(p-q) + {\cal L} (x,p|y,q)\right]^{-1} .
\end{align}
Because the linearized Boltzmann equation is causal and purely
relaxational, the two-point function in \eqref{eq:10} is
retarded. This implies that it is possible to extract the collision integral of the linearized Boltzmann equation directly from the analytic structure of the retarded Green's function $G^{\rho\rho}_R(x,p|y,q)$. As a result, the eigenvalues of the Boltzmann equation $\xi_n$ are also the locations of the poles of  $G^{\rho\rho}_R$. This establishes a direct connection  between weakly coupled quantum field theory and quantum kinetic theory. From the definition of $\rho(k,p)$, Eq. \eqref{eq:10} can be
expressed in terms of the connected\footnote{The disconnected part
  gives a product of the equilibrium one-point functions $\langle
  \rho\rangle_{\beta}$.}
 Schwinger-Keldysh (SK) four-point
functions (see Ref. \cite{Wang:2002nba}) of the microscopic fields $G^{\rho\rho}_R(k,p;\ell,q) = -  G^{\Phi^2\Phi^2}_{1111}+G^{\Phi^2\Phi^2}_{1122}$, where
\begin{align}
  \label{eq:13}
G^{\Phi^2\Phi^2}_{1122}=&\,\,i \left\langle \mathrm{Tr} \left[\Phi_1(p+k/2)\Phi_1(-p+k/2) \right] \right.\nn
&\times \left. \mathrm{Tr} \left[\Phi_2(q+\ell/2)\Phi_2(-q+\ell/2) \right]\right\rangle_{SK} ,
\end{align}
and similarly for $G_{1111}^{\Phi^2\Phi^2}$. Here, $\Phi_{1,2}$ denote the doubled fields on the forward and backward
contours of the SK path integral, respectively. In translationally invariant systems, \mbox{$\ell=-k$}. It is convenient to introduce the Keldysh basis,
\mbox{$\Phi_a = \Phi_1-\Phi_2$} and $\Phi_r \!=\!\!
\frac{1}{2}(\Phi_1+\Phi_2)$. Then $G^{\rho\rho}_R$ is a linear
combination of 16 four-point functions
$G_{\alpha_1\alpha_2\alpha_3\alpha_4} =
i2^{n_{r_{\alpha_i}}}\langle\Phi_{\alpha_1}\Phi_{\alpha_2}\Phi_{\alpha_3}\Phi_{\alpha_4}\rangle$
with $\alpha_i=\{a,r\}$ and $n_{r_{\alpha_i}}$ counting the number of
$\alpha_i$ indices equal to $r$. In the limit of small frequency and
momenta, $\omega \equiv k^0 \to 0$ and $\vec k \to 0$, however,
it is only a single one of these four-point
functions that contributes to the final expression \cite{Wang:2002nba,Czajka:2017bod,Extended}:
\begin{align}
&\lim_{k\rar0} G^{\rho\rho}_R(p,q|k) = - \lim_{k\rar0} \frac{\beta
  k^0}{2} \, \CN(p^0) G^*_{aarr}(p,q|k) \nn
&= - \lim_{k\rar0} \frac{\beta k^0}{4} \, \CN(p^0) \,\CN(q^0) \,
  \langle f(p,k)f(q,-k)\rangle \,,
\label{eq:Identity}
\end{align}
where $\CN(p^0) =  n(p^0)\left(1 + n(p^0)\right)$. The exact four-point
function $G^*_{aarr}(p,q|k)$ obeys a system of Bethe-Salpeter
equations (BSEs) that nevertheless still couples all 16
$G_{\alpha_1\alpha_2\alpha_3\alpha_4}$. However, it turns out that in
the limit of small $\omega$ and $\vec k$, $G^*_{aarr}$ decouples and is governed by a single BSE \cite{Wang:2002nba,Extended}:
\begin{align}\label{BSE-Boltzmann1}
G^*_{aarr}(p,q|k)= & \,\, \Delta_{ra}(p+k)\Delta_{ar}(p)\biggr[ i(2\pi)^4\delta^4(p-q)N^2  \nn
& -\int_l \mathcal{K}(p,l)G^*_{aarr}(l,q|k)\biggr]\,,
\end{align}
where $\Delta_{\alpha_1\alpha_2} = -i\,
2^{n_{r_{\alpha_i}}}\langle\Phi_{\alpha_1}\Phi_{\alpha_2}\rangle$
is the Schwinger-Keldysh two-point function and $\mathcal{K}(p,\ell) =
d\Sigma_{p\rar l}/ \CN(p^0)$, with $d\Sigma_{p\rar l}$ 
the transition probability of an off-shell particle
with energy-momentum  $(p^0,\vec{p})$ scattering of the thermal bath
to an off-shell particle with energy-momentum $(l^0,\vec
l)$.\footnote{$\mathcal{K}(p,\ell) =- \frac{\sinh(\beta
    p^0/2)}{\sinh(\beta\ell^0/2)}R(l-p)$ where $R(l-p)$ is the rung
  function computed in \cite{Stanford:2015owe}. Note that the $R(l-p)$
  in \cite{Stanford:2015owe} is not the same as $R^\wedge$ or $R^\vee$
  used here.} Defining $G^*_{aarr}(p|k)= \int_q G^*_{aarr}(p,q|k)$, Eq. \eqref{BSE-Boltzmann1} reduces to
\begin{align}
 G^*_{aarr}(p|k)=& \,\, \Delta_{ra}(p+k)\Delta_{ar}(p)\biggr [i N^2  \nn 
 & -   \int_l  \mathcal{K}(p,l)G^*_{aarr}(l|k)\biggr]\, .
\end{align}
The product $\Delta_{ra}(p+k)\Delta_{ar}(p)$ has four poles with imaginary parts $\pm i\Gamma_\vec{p}$. However, as $k \rightarrow 0$, only a contribution from two poles remains. This pinching pole approximation, ubiquitous in the study of hydrodynamic transport coefficients and spectra of finite temperature quantum field theories \cite{Jeon:1994if,Wang:2002nba}, gives
\begin{align}
\label{eq:BSE}
G^*_{aarr}(p|k) = \! \frac{\pi}{E_{\mathbf{p}}}\frac{\delta(p_0^2-E_{\mathbf{p}}^2)}{-i\omega+2\Gamma_{\mathbf{p}}}\biggr[i
  N^2 \! -\int_l \mathcal{K}(p,l) G^*_{aarr}(l|k)\biggr] .
\end{align}

To find the solution of the integral equation \eqref{eq:BSE}, we make the ansatz whereby $G^*_{aarr}(p|k)$ is supported on-shell: 
\begin{align}
G^*_{aarr}(p|k) = \delta(p_0^2-E^2_{\mathbf{p}})G^{ff}(\mathbf{p}|k).
\end{align} 
Hence,
\begin{align}\label{eq:EvenAnsatz}
&(-i\omega+2\Gamma_{\mathbf{p}})G^{ff}(\vec{p}|k)=\frac{i \pi N^2}{E_\vec{p}} -\int_{\vec l} \, \frac{1}{2E_{\vec{p}}} \nn
&\times  \left[     \mathcal{K}(\vec{p},E_\vec{p}|\vec{\ell},E_\vec{\ell})+\mathcal{K}(\vec{p},E_\vec{p}|\vec{\ell},-E_\vec{\ell})\right]  G^{ff}(\mathbf{l}|k) \,.
\end{align}
It can be shown that 
\begin{align}
\frac{1}{2E_{\vec{p}}}\mathcal{K}(\vec{p},E_\vec{p}|\vec{\ell},E_\vec{\ell})=-R^\wedge(\vec{p},\vec{l}) 
\end{align}
and \cite{Jeon:1994if,Wang:2002nba,Extended}
\begin{align}
\frac{1}{2E_{\vec{p}}}{\mathcal{K}(\vec{p},E_\vec{p}|\vec{l},-E_\vec{l})}=R^\vee(\vec{p},\vec{l})-
2\Gamma_{\mathbf{p}}\delta(\vec{p}-\vec{l}) \,.
\end{align} 
Thus, Eq. \eqref{eq:EvenAnsatz} is solved by
\begin{equation}
\!\! G^{ff}(\vec{p}|k)=\frac{i\pi N^2}{E_{\vec{p}}} \, \frac{1}{-i\omega -\int_{\vec l} \left[ R^\wedge(\vec{p},\vec{k})-R^\vee(\vec{p},\vec{k})\right]} \,.
\end{equation}
Hence, the spectrum of $G^{ff}(\vec{p}|k^0=\omega,\vec{k}=0)$ equals the spectrum of the one-particle distribution $f(t,\vec{p})$ determined by the linearized Boltzmann equation \eqref{eq:2}. 

The derivation of the kinetic equation \eqref{eq:fEXQ} for quantum
chaos from the OTOC now follows from an analogous line of arguments. The OTOC, 
\begin{align}
  \label{eq:11}
  C(t) =& - i\int_{k} e^{-ikt} \int\displaylimits_{p,q} \left\langle [\Phi_{ab}(p+k),\Phi_{a'b'}^{\dagger}(-q-k)] \right. \nn
  &\left. \times  [\Phi_{ab}^{\dagger}(-p),\Phi_{a'b'}(q)] \right\rangle \,,
\end{align}
is a four-point function, which, as shown in \cite{Stanford:2015owe}, also obeys a BSE in the limit of $\omega \to 0$. Indicating with $\cG_{\scriptscriptstyle OTOC}(p,q|k,p+q-k)$ the term inside the integrals in Eq. \eqref{eq:11}, i.e. $C(t) \equiv \int_k e^{-ikt} \int_{p,q} \cG_{\scriptscriptstyle OTOC}(p,q|k,p+q-k)$, we define 
\begin{align}
\widetilde{\cG}(p|k) = \int_q  \cG_{\scriptscriptstyle OTOC}(p,q|k,p+q-k) \,.
\end{align}
The correlator $\widetilde{\cG}(p|k) $ then obeys the following integral equation: 
\begin{align}\label{eq:BSE2}
\widetilde{\cG}(p|k) =&\,
  \frac{\pi}{E_{\mathbf{p}}}\frac{\delta(p_0^2-E_{\mathbf{p}}^2)}{-i\omega
  + 2\Gamma_{\mathbf{p}}} \biggr[iN^2 \nn
&- \int \frac{d^4\ell}{(2\pi)^4} \frac{\sinh(\beta \ell_0/2)}{\sinh(\beta p_0/2)} \mathcal{K}(p,l)\widetilde{\cG}(\ell|k)\biggr].
\end{align}
Eq. \eqref{eq:BSE2} agrees with the result found in \cite{Stanford:2015owe}, even though it is expressed here with different notation. The advantage of writing $\widetilde{\cG}(p|k) $ as in \eqref{eq:BSE2} is that it makes transparent the similarities between $\widetilde{\cG}(p|k) $ and $G^*_{aarr}(p|k)$ from Eq. \eqref{eq:BSE}, which governs transport. A priori, there is no reason to expect $\widetilde{\cG}(p|k)$ and $G^*_{aarr}(p|k)$ to be related. Nevertheless, by comparing \eqref{eq:BSE} with \eqref{eq:BSE2}, it is clear that in this calculation, the only difference between the two BSE equations is the factor $\frac{\sinh(\beta \ell_0/2)}{\sinh(\beta p_0/2)}$ appearing in the measure of the kernel of \eqref{eq:BSE2}. As we will see in Section \ref{sec:Results}, this factor is crucial for the fact that, while related, the spectra of $G^*_{aarr}(p|k)$ and $\widetilde{\cG}(p|k)$ are distinct: the spectrum of $G^*_{aarr}(p|k)$ only possesses relaxational modes while $\widetilde{\cG}(p|k)$ exhibits exponentially growing modes which can be associated with many-body quantum chaos. 

To find a solution of Eq. \eqref{eq:BSE2}, as in the case of Eq. \eqref{eq:BSE}, we again introduce an on-shell ansatz $\widetilde{\cG}(p|k) = \delta(p_0^2-E^2_{\mathbf{p}})G^{\mathsf{ff}}(\vec{p}|k)$. This gives
\begin{align}\label{eq:OTOC}
&(-i\omega+2\Gamma_{\mathbf{p}})G^{\mathsf{ff}}(\vec{p}|k)=\frac{i \pi N^2}{E_\vec{p}}  -\int_{\vec l} \, \frac{\sinh(\beta E_\vec{l}/2)}{\sinh(\beta E_\vec{p}/2)}\frac{1}{2E_{\vec{p}}}      \nn
&\times \left(\mathcal{K}(\vec{p},E_\vec{p}|\vec{l},E_\vec{l})-\mathcal{K}(\vec{p},E_\vec{p}|\vec{l},-E_\vec{l})\right)G^{\mathsf{ff}}(\mathbf{l}|k) ,
\end{align}
where one of the signs in front of $\mathcal{K}$ is now reversed due to the fact that factor $\frac{\sinh(\beta \ell_0/2)}{\sinh(\beta p_0/2)}$ in the measure is an odd function of energy. Thus, the spectrum of $G^{\mathsf{ff}}(\mathbf{l}|k)$, and hence, of the OTOC,
equals the spectrum of the following kinetic equation 
\begin{align}\label{eq:kineticOTOC}
&\partial_t f_{\scriptscriptstyle OTOC} (t,\vec p) = \int_{\vec{l}}\frac{\sinh(\beta E_\vec{l}/2)}{\sinh(\beta E_\vec{p}/2)}\nn
&\times \left[ R^\wedge(\vec{p},\vec{l})+R^\vee(\vec{p},\vec{l}) -4\Gamma_{\vec{p}}\delta(\vec{p}-\vec{l})\right] f_{\scriptscriptstyle OTOC}(t,\mathbf{l}) \,,
\end{align}
which precisely matches with the kinetic equation for the OTOC put forward in Eq. \eqref{eq:fEXQ}, with
$\CE(E_\vec{p}) = 1 / \sinh(\beta E_\vec{p} /2)$, or $\lim_{\beta\to0}
\CE(E_\vec{p})/\CE(E_\vec{l}) = E_\vec{l} / E_\vec{p}$. As noted
there, this spectrum of Eq. \eqref{eq:fEXQ} is in fact independent of
$\CE(E)$ as long as the function $\CE$ is odd.

\section{Results and discussion}\label{sec:Results}

\begin{figure}[h]
\centering
\includegraphics[width=0.86\textwidth]{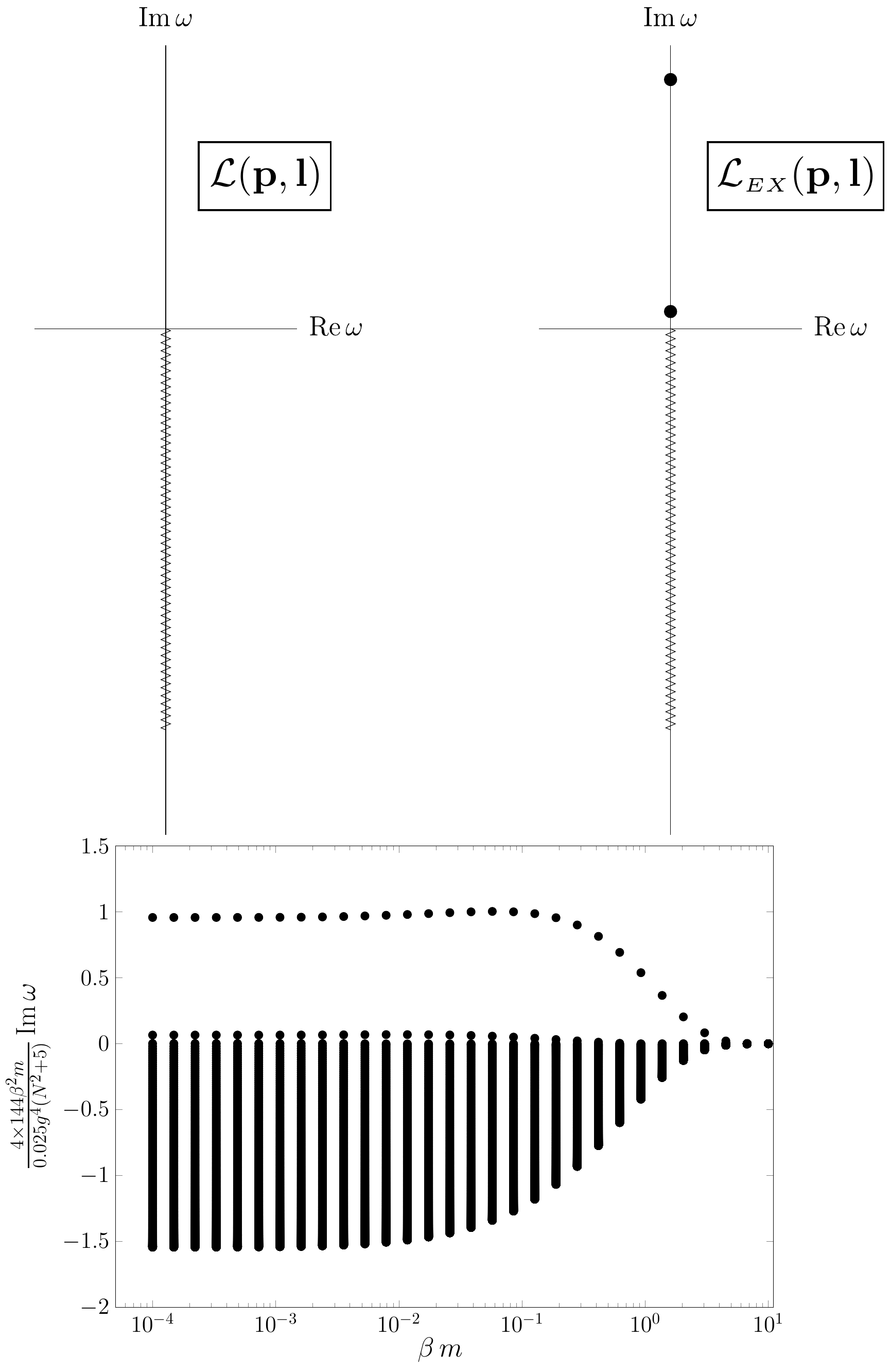}
\caption{The spectra of the kernel $\CL(\vec{p},\vec{l})$ for the linearized Boltzmann
  equation (and also of $\left\la T^{xy}(k_z), T^{xy}(-k_z) \right\ra_R$, cf. Eq. \eqref{Txyxy}) (top left) and of the kernel $\CL_{EX}(\vec{p},\vec{l})$ for the kinetic equation for the OTOC (top right) are plotted over the
  complex $\omega$ plane and in the limit of $\beta m \to 0$. In the lower half of the complex $\omega$ plane, there is a dense sequence of numerically obtained poles. In both spectra, these poles are believed to be the signature of a branch cut. See \cite{Moore:2018mma} and also
  \cite{Romatschke:2015gic,Grozdanov:2016vgg,Kurkela:2017xis,Solana:2018pbk}.  In the upper half of the complex $\omega$ plane, only the kernel $\CL_{\scriptscriptstyle EX}(\vec{p},\vec{l})$ has distinct poles which are identified with the Lyapunov exponents, as explained below equation \eqref{eq:label}. The dependence of these two Lyapunov exponents and the branch cuts on $\beta m$ is depicted in the inlay (bottom). For large values of $\beta m$, the Lyapunov exponents decay exponentially. The plots are obtained by diagonalizing the kernels of the integral equations \eqref{BSE-Boltzmann1} and \eqref{eq:kineticOTOC} after a discretization with $N=1000$ grid points on the domain $p\in[m/N, N\times m]$. The discretization is not uniform. This is done in order for the diagonalization to appropriately account for the contributions of both the soft momenta and collinear momenta $\vec{p}\approx \vec{l}$, which are not negligible even when both $\vec{p}$ and $\vec{l}$ are  large \cite{Jeon:1994if,Extended}. The finite size of the branch cuts, i.e. its end point for large Im($\omega$), is related to finite domain of the discretization procedure.
}
\label{fig:Spect}
\end{figure}

In addition to greatly simplifying the computation
of chaotic behavior in dilute weakly interacting systems and providing a physical
picture for the meaning of many-body chaos, the gross energy exchange
kinetic equation recasting of the OTOC makes it conspicuously clear how
in such systems scrambling (or ergodicity) and transport are governed
by the same physics \cite{Grozdanov:2017ajz}. The kernel of the
kinetic equation in both cases is the 2-to-2 scattering
cross-section. Nevertheless, the equations for $f_{\scriptscriptstyle
  OTOC}$, or equivalently, $f_{\scriptscriptstyle
  EX}$, and $f$ are subtly different, which allows for the crucial
qualitative difference: a chaotic, Lyapunov-type divergent growth of $f_{\scriptscriptstyle
  EX}$ versus damped relaxation of $f$. Their spectra at $\vec k =0$ and small $\omega$ are presented in Figure \ref{fig:Spect}. As already noted below Eq. \eqref{eq:BSE2}, the two \emph{off-shell} late-time BSEs \eqref{eq:BSE} and \eqref{eq:BSE2} are the same upon performing the following identification: 
\begin{align}
\widetilde{\cal G}(p|k) =G^{\ast}_{aarr}(p|k)/\sinh(\beta p^0/2).
\end{align} 
The most general solution to this BSE thus includes the information
about chaos and transport. However, the divergent modes (in time)
  of the OTOC are projected out by the on-shell condition and thus do not contribute to the correlators that compute transport. For example, the shear viscosity $\eta$ can be inferred from the following retarded correlator (see e.g. \cite{Jeon:1994if}): 
\begin{align}\label{Txyxy}
\left\la T^{xy}(k), T^{xy}(-k) \right\ra_R  = \int_{p,q} p_x p_y q_x q_y\, G^{\rho\rho}_R(k |p,q) \,,
\end{align}
where $k = (\omega,0,0,k_z)$. The integrals over $p$ and $q$, together with the on-shell condition, project out the odd modes in $p^0$ which govern chaos, and transport is only sensitive to the even, stable modes \cite{Moore:2018mma}. 

The fact that, when off shell, the BSEs \eqref{eq:BSE} and \eqref{eq:BSE2} can be mapped onto each other is by itself a highly non-trivial result which opens several questions. In particular, this observation seems to indicate that in some cases, the information about scrambling and ergodicity, which has so far been believed to be accessible only by studying a modified, extended SK contour and OTOCs, can instead be addressed by a suitable analysis of the analytic properties of correlation functions on the standard SK contour. How our result implies such new analytical properties, remains to be discovered. We remark, however, that studies in (holographic) strongly
coupled theories uncovered precisely this type of a relation between hydrodynamic transport at an analytically continued imaginary momentum and chaos. In particular, as we discovered in \cite{Grozdanov:2017ajz}, chaos is encoded in a vanishing residue (``pole-skipping'') of the retarded energy density two-point function, which tightly constrains the behavior of the dispersion relation of longitudinal (sound) hydrodynamic excitations. The same imprint of chaos on properties of transport was later also observed in a proposed effective (hydrodynamic) field theory of chaos \cite{Blake:2017ris}. Despite the fact that it is at present unknown how general pole-skipping is and whether other related analytic signatures of chaos in observables that characterize transport exist, it may be possible that properties of many-body quantum chaos in dilute weakly coupled theories are also uncoverable from transport, as in strongly coupled theories \cite{Grozdanov:2017ajz}. We defer these questions to future works.

The kinetic equation for many-body chaos, that we have derived here, also gives
concrete form to past attempts to do so, which were based on a
phenomenological ansatz that one should count additively the number of
collisions \cite{PhysRevLett.80.2035,1999chao.dyn..9034V}. In essence,
that is also what our gross exchange equation does. The exponential divergence can thus be understood as a front propagation into unstable states \cite{Saarloos:2003}. This analogy was already noted in \cite{Aleiner:2016eni} who derived a kinetic equation for chaos from the Dyson equation for the $4\times 4$ matrix of the
four-contour SK Green's functions. By our arguments above that relate
the poles of the OTOC to a dynamical equation for
$f_{\scriptscriptstyle OTOC}$, the resulting equations in
\cite{Aleiner:2016eni} should contain a decoupled subsector that is
equivalent to the kinetic equation derived here.

Finally, we wish to note that the small parameter that sets the Ehrenfest time and controls the regime of exponential growth in the OTOC in all these systems is the perturbative small 't Hooft coupling $\lambda=g^2 N$. The BSE from which the kinetic equation is derived is formally equivalent to a differential equation of the type
\begin{align}
  \label{eq:14}
  \left(\frac{d}{dt}-g^4 N^2 L\right) f = N^2 \,.
\end{align}
This is solved by 
\begin{align}
f = -\frac{1}{g^4 L} + c_0 \,e^{g^4 N^2 Lt}.
\end{align} 
The Ehrenfest (or scrambling) time, where the exponential becomes of order of the constant term, is therefore \begin{align}
t_{scr} = \frac{1}{g^4 N^2 L} \ln(1/g^4 Lc_0).
\end{align} 
For small $g^2$, this can be an appreciable
timescale for any value of $N$, and there is no need for a large $N$ number of species.

\acknowledgements

We are very grateful to Pavel Friedrich, Hong Liu, Guy Moore and Toma\v{z} Prosen for discussions and Wim van Saarloos for calling our attention to his work \cite{Saarloos:2003}. This research was supported in part by a VICI award of the Netherlands Organization for Scientific Research (NWO), by the Netherlands Organization for Scientific Research/Ministry of Science and Education (NWO/OCW), and by the Foundation for Research into Fundamental Matter (FOM). S. G. is supported by the U.S. Department of Energy under grant Contract Number DE-SC0011090.

\bibliographystyle{apsrev4-1}
\bibliography{references}

\appendix

\section{Diagrammatic expansion of $|{\cal T}_{12\rar 34}|^2$ in the theory of $N\times N$ Hermitian matrix scalars}\label{App}

Here, we present the diagrammatic expansion and the relevant
combinatorial factors for each of the diagrams that enter into the
2-to-2 transition amplitude $|{\cal T}_{12\rar 34}|$ in the 
theory of $N\times N$ Hermitian matrix scalars (9). The square of the 2-to-2
transition amplitude, $|{\cal T}_{12\rar 34}|^2$, is the square of the
amputated connected four-point function. At lowest non-trivial order:

\begin{center}
\begin{tikzpicture}[scale=0.5][h]
\draw[dashed] (0,-3)--(0,3);
\draw[thick,decoration={markings, mark=at position 0.5 with {\arrow[line width=0.5mm]{<}}},postaction={decorate}] (0,2) -- (2,0);
\draw[thick,decoration={markings, mark=at position 0.5 with {\arrow[line width=0.5mm]{<}}},postaction={decorate}] (2,0) -- (4,-2);
\draw[thick,decoration={markings, mark=at position 0.5 with {\arrow[line width=0.5mm]{<}}},postaction={decorate}] (0,-2) -- (2,0);
\draw[thick,decoration={markings, mark=at position 0.5 with {\arrow[line width=0.5mm]{<}}},postaction={decorate}] (2,0) -- (4,2);
\begin{scope}[xscale=-1]
\draw[thick,decoration={markings, mark=at position 0.5 with {\arrow[line width=0.5mm]{<}}},postaction={decorate}] (0,2) -- (2,0);
\draw[thick,decoration={markings, mark=at position 0.5 with {\arrow[line width=0.5mm]{<}}},postaction={decorate}] (2,0) -- (4,-2);
\draw[thick,decoration={markings, mark=at position 0.5 with {\arrow[line width=0.5mm]{<}}},postaction={decorate}] (0,-2) -- (2,0);
\draw[thick,decoration={markings, mark=at position 0.5 with {\arrow[line width=0.5mm]{<}}},postaction={decorate}] (2,0) -- (4,2);
\end{scope}
\end{tikzpicture}
\end{center}
For $N=1$ the theory is just scalar $\phi^4$ theory and the answer is straightforward: $|{\cal T}_{12\rar
  34}|^2 = g^4$.

For $N>1$ theory, the actual amplitude we wish to compute is additionally traced
over the external indices, since, 
\begin{align}
  \label{eq:app:11}
  C(t) =& - i\int_{k} e^{-ikt} \int\displaylimits_{p,q} \left\langle [\Phi_{ab}(p+k),\Phi_{a'b'}^{\dagger}(-q-k)] \right. \nn
  &\left. \times  [\Phi_{ab}^{\dagger}(-p),\Phi_{a'b'}(q)] \right\rangle \,,
\end{align}
The way that the matrix indices need to be contracted is across the
cut. An easy way to see this from
the free non-interacting result: $C(t)_{g^2=0} = G^{ab,cd}_{R}
G_{cd,ab;R}$. Graphically, 

\begin{center}
\begin{tikzpicture}[scale=0.5][h]
\draw[dashed] (0,-3)--(0,3);
\draw[thick,decoration={markings, mark=at position 0.5 with {\arrow[line width=0.5mm]{<}}},postaction={decorate}] (0,2) -- (2,0);
\draw[thick,decoration={markings, mark=at position 0.5 with {\arrow[line width=0.5mm]{<}}},postaction={decorate}] (2,0) -- (4,-2);
\draw[thick,decoration={markings, mark=at position 0.5 with {\arrow[line width=0.5mm]{<}}},postaction={decorate}] (0,-2) -- (2,0);
\draw[thick,decoration={markings, mark=at position 0.5 with {\arrow[line width=0.5mm]{<}}},postaction={decorate}] (2,0) -- (4,2);
\begin{scope}[xscale=-1]
\draw[thick,decoration={markings, mark=at position 0.5 with {\arrow[line width=0.5mm]{<}}},postaction={decorate}] (0,2) -- (2,0);
\draw[thick,decoration={markings, mark=at position 0.5 with {\arrow[line width=0.5mm]{<}}},postaction={decorate}] (2,0) -- (4,-2);
\draw[thick,decoration={markings, mark=at position 0.5 with {\arrow[line width=0.5mm]{<}}},postaction={decorate}] (0,-2) -- (2,0);
\draw[thick,decoration={markings, mark=at position 0.5 with {\arrow[line width=0.5mm]{<}}},postaction={decorate}] (2,0) -- (4,2);
\end{scope}
\draw[thick,gray] (-4.2,2.2) to
[out=135,in=45] (4.2,2.2);
\begin{scope}[yscale=-1]
\draw[thick,gray] (-4.2,2.2) to
[out=135,in=45] (4.2,2.2);
\end{scope}
\end{tikzpicture}
\end{center}
Above the arrows denote momentum-flow. We are interested in the way
the weight changes as a function of $N$.

To find this answer, we use that Hermitian matrices span the adjoint
of $U(N)$. Following 't Hooft, one can then use double line notation in terms of
fundamental $N$-``charges''. Using this double line notation, the
vertex equals.
\begin{center}
\begin{tikzpicture}[scale=0.5][h]
\draw[doubleline] (0,2) -- (1.9,0.1);
\draw[doubleline] (2.1,-.1) -- (4,-2);
\draw[doubleline] (0,-2) -- (1.9,-.1);
\draw[doubleline] (2.1,.1) -- (4,2);
\end{tikzpicture}
\end{center}
One needs to connect the two vertices across the cut, and then
contract, i.e. trace over the external indices, in all possible ways.
We will do so step-wise.

Consider first the transition probability. Connecting the first leg
across the cut is unambiguous, i.e., each possible choice gives the
same answer:
\begin{center}
\begin{tikzpicture}[scale=0.5][h]
\node at (-1,0) {$|{\cal T}_{12\rar 34}|^2=~~~~~~~~~ $}; 
\draw (0,-3)--(0,3);
\begin{scope}[xshift=1cm]
\draw[doubleline] (-1,2).. controls (0,2) .. (1.9,0.1);
\draw[doubleline] (2.1,-.1) -- (4,-2);
\draw[doubleline] (0,-2) -- (1.9,-.1);
\draw[doubleline] (2.1,.1) -- (4,2);
\end{scope}
\draw (6,-3)--(6,3);
\node[above right] at (6,3) {$\Huge 2$};
\end{tikzpicture} ~
\end{center}
Contracting the next line, however, gives rise to in-equivalent
possibilities, each with the same weight $w$.
They are
\begin{widetext}
\begin{center}
\begin{tikzpicture}[scale=0.5][h]
\node at (-1,0) {$|{\cal T}_{12\rar 34}|^2= w^2~~~~~~~~~~~~ $}; 
\draw (0,-3)--(0,3);
\begin{scope}[xshift=1cm]
\draw[doubleline] (-1,2).. controls (0,2) .. (1.9,0.1);
\draw[doubleline] (2.1,-.1) -- (4,-2);
\draw[doubleline] (-1,-2).. controls (0,-2) .. (1.9,-.1);
\draw[doubleline] (2.1,.1) -- (4,2);
\end{scope}
\begin{scope}[xshift=6.5cm]
\draw[doubleline] (0,2).. controls (2,2) .. (2,0.1);
\draw[doubleline] (2.1,0) -- (3,0);
\draw[doubleline] (0,-2).. controls (2,-2) ..  (2,-.1);
\draw[doubleline] (1,0) -- (1.9,0);
\end{scope}
\begin{scope}[xshift=11cm]
\draw[doubleline] (0,2).. controls (4,2) .. (2,0.1);
\draw[doubleline] (1,1) -- (1.9,0.1);
\draw[doubleline] (0,-2).. controls (4,-2) ..  (2,-0.1);
\draw[doubleline] (1,-1) -- (1.9,-.1);
\end{scope}
\node[right] at (5,0) {$\Huge +$};
\node[right] at (10,0) {$\Huge +$};
\draw (16,-3)--(16,3);
\node[above right] at (16,3) {$\Huge 2$};
\end{tikzpicture} 
\end{center}
\end{widetext}
Now, multiplying out the various combinations, each of the six
independent combinations can be contracted in two ways
over the external indices. As a result, we obtain the following set of twelve independent diagrams.

{\bf Diagram 1} with weight $N^4$ and multiplicity $1$:

\begin{center}
\begin{tikzpicture}[scale=0.4][h]
  \begin{scope}
\draw[doubleline] (-1,2).. controls (0,2) .. (1.9,0.1);
\draw[doubleline] (2.1,-.1) -- (4,-2);
\draw[doubleline] (-1,-2).. controls (0,-2) .. (1.9,-.1);
\draw[doubleline] (2.1,.1) -- (4,2);
\end{scope}
\begin{scope}[xscale=-1,xshift=2cm]
\draw[doubleline] (-1,2).. controls (0,2) .. (1.9,0.1);
\draw[doubleline] (2.1,-.1) -- (4,-2);
\draw[doubleline] (-1,-2).. controls (0,-2) .. (1.9,-.1);
\draw[doubleline] (2.1,.1) -- (4,2);
\end{scope}
\draw[doubleline,gray] (4.2,2.2) to [out=45, in =135] (-6.2,2.2);
\begin{scope}[yscale=-1]
\draw[doubleline,gray] (4.2,2.2) to [out=45, in =135] (-6.2,2.2);
\end{scope}
\end{tikzpicture}
\end{center}

{\bf Diagram 2} with weight $N^2$ and multiplicity $1$:

\begin{center}
\begin{tikzpicture}[scale=0.4][h]
\draw[doubleline,gray] (-6.2,2.2) to [out=135, in =135] (-4.2,3.2) to (2.2,-3.2) to [out=315, in=315] (4.2,-2.2);
\begin{scope}[yscale=-1]
\draw[doubleline,gray] (-6.2,2.2) to [out=135, in =135] (-4.2,3.2) to (2.2,-3.2) to [out=315, in=315] (4.2,-2.2);
\end{scope}
  \begin{scope}
\draw[doubleline] (-1,2).. controls (0,2) .. (1.9,0.1);
\draw[doubleline] (2.1,-.1) -- (4,-2);
\draw[doubleline] (-1,-2).. controls (0,-2) .. (1.9,-.1);
\draw[doubleline] (2.1,.1) -- (4,2);
\end{scope}
\begin{scope}[xscale=-1,xshift=2cm]
\draw[doubleline] (-1,2).. controls (0,2) .. (1.9,0.1);
\draw[doubleline] (2.1,-.1) -- (4,-2);
\draw[doubleline] (-1,-2).. controls (0,-2) .. (1.9,-.1);
\draw[doubleline] (2.1,.1) -- (4,2);
\end{scope}
\end{tikzpicture}
\end{center}

{\bf Diagram 3} with weight $N^2$ and multiplicity $2$ (a crossterm diagram):
\begin{center}
\begin{tikzpicture}[scale=0.4][h]
\draw[doubleline,gray] (-5.2,2.2) to [out=135, in =135] (-4.2,3.2) to (-.2,.2) to [out=315, in=180] (1,0);
  \begin{scope}
\draw[doubleline] (0,2).. controls (2,2) .. (2,0.1);
\draw[doubleline] (2.1,0) -- (3,0);
\draw[doubleline] (0,-2).. controls (2,-2) ..  (2,-.1);
\draw[doubleline] (1.2,0) -- (1.9,0);
\end{scope}
\begin{scope}[xscale=-1,xshift=1cm]
\draw[doubleline] (-1,2).. controls (0,2) .. (1.9,0.1);
\draw[doubleline] (2.1,-.1) -- (4,-2);
\draw[doubleline] (-1,-2).. controls (0,-2) .. (1.9,-.1);
\draw[doubleline] (2.1,.1) -- (4,2);
\end{scope}
\draw[doubleline,gray] (-5.2,-2.2) to [out=215, in =0] (-2,-3) -- (1,-3) to [out=0,in=0] (3,0);
\end{tikzpicture}
\end{center}

{\bf Diagram 4} with weight $N^2$ and multiplicity $2$ (a crossterm diagram). It equals Diagram 3 mirrored across the horizontal axis:

\begin{center}
\begin{tikzpicture}[scale=0.4][h]
\begin{scope}[yscale=-1]
\draw[doubleline,gray] (-5.2,2.2) to [out=135, in =135] (-4.2,3.2) to (-.2,.2) to [out=315, in=180] (1,0);
  \begin{scope}
\draw[doubleline] (0,2).. controls (2,2) .. (2,0.1);
\draw[doubleline] (2.1,0) -- (3,0);
\draw[doubleline] (0,-2).. controls (2,-2) ..  (2,-.1);
\draw[doubleline] (1.2,0) -- (1.9,0);
\end{scope}
\begin{scope}[xscale=-1,xshift=1cm]
\draw[doubleline] (-1,2).. controls (0,2) .. (1.9,0.1);
\draw[doubleline] (2.1,-.1) -- (4,-2);
\draw[doubleline] (-1,-2).. controls (0,-2) .. (1.9,-.1);
\draw[doubleline] (2.1,.1) -- (4,2);
\end{scope}
\draw[doubleline,gray] (-5.2,-2.2) to [out=215, in =0] (-2,-3) -- (1,-3) to [out=0,in=0] (3,0);
\end{scope}
\end{tikzpicture}
\end{center}

{\bf Diagram 5} with weight $N^2$ and multiplicity $2$ (a crossterm diagram):

\begin{center}
\begin{tikzpicture}[scale=0.4][h]
\draw[doubleline,gray] (-5.2,-2.2) to [out=215, in =0] (-2,-3) to [out=0, in=215] (0.9,-1.1); 
\begin{scope}[yscale=-1]
\draw[doubleline,gray] (-5.2,-2.2) to [out=215, in =0] (-2,-3) to [out=0, in=215] (0.9,-1.1); 
\end{scope}
\begin{scope}
\draw[doubleline] (0,2).. controls (4,2) .. (2,0.1);
\draw[doubleline] (1,1) -- (1.9,0.1);
\draw[doubleline] (0,-2).. controls (4,-2) ..  (2,-0.1);
\draw[doubleline] (1,-1) -- (1.9,-.1);
\end{scope}
\begin{scope}[xscale=-1,xshift=1cm]
\draw[doubleline] (-1,2).. controls (0,2) .. (1.9,0.1);
\draw[doubleline] (2.1,-.1) -- (4,-2);
\draw[doubleline] (-1,-2).. controls (0,-2) .. (1.9,-.1);
\draw[doubleline] (2.1,.1) -- (4,2);
\end{scope}
\end{tikzpicture}
\end{center}

{\bf Diagram 6} with weight $N^2$ and multiplicity $2$ (a crossterm diagram):

\begin{center}
\begin{tikzpicture}[scale=0.4][h]
\draw[doubleline,gray] (-5.2,-2.2) to [out=215, in =215] (-4.5,-3) to (-0.5,1) to [out=45, in=135] (0.9,1.1); 
\begin{scope}[yscale=-1]
\draw[doubleline,gray] (-5.2,-2.2) to [out=215, in =215] (-4.5,-3) to (-0.5,1) to [out=45, in=135] (0.9,1.1); 
\end{scope}
\begin{scope}
\draw[doubleline] (0,2).. controls (4,2) .. (2,0.1);
\draw[doubleline] (1,1) -- (1.9,0.1);
\draw[doubleline] (0,-2).. controls (4,-2) ..  (2,-0.1);
\draw[doubleline] (1,-1) -- (1.9,-.1);
\end{scope}
\begin{scope}[xscale=-1,xshift=1cm]
\draw[doubleline] (-1,2).. controls (0,2) .. (1.9,0.1);
\draw[doubleline] (2.1,-.1) -- (4,-2);
\draw[doubleline] (-1,-2).. controls (0,-2) .. (1.9,-.1);
\draw[doubleline] (2.1,.1) -- (4,2);
\end{scope}
\end{tikzpicture}
\end{center}

{\bf Diagram 7} with weight $N^4$ and multiplicity $1$:

\begin{center}
\begin{tikzpicture}[scale=0.4][h]
\draw[doubleline,gray] (-1,0)-- (1,0);
\draw[doubleline,gray] (-3.1,0) to [out=180,in=180] (-1,-3) -- (1,-3) to [out=0,in=0] (3.1,0);
  \begin{scope}
\draw[doubleline] (0,2).. controls (2,2) .. (2,0.1);
\draw[doubleline] (2.1,0) -- (3,0);
\draw[doubleline] (0,-2).. controls (2,-2) ..  (2,-.1);
\draw[doubleline] (1.2,0) -- (1.9,0);
\end{scope}
\begin{scope}[xscale=-1,xshift=0cm]
\draw[doubleline] (0,2).. controls (2,2) .. (2,0.1);
\draw[doubleline] (2.1,0) -- (3,0);
\draw[doubleline] (0,-2).. controls (2,-2) ..  (2,-.1);
\draw[doubleline] (1.2,0) -- (1.9,0);
\end{scope}
\end{tikzpicture}
\end{center}

{\bf Diagram 8} with weight $N^2$ and multiplicity $1$:

\begin{center}
\begin{tikzpicture}[scale=0.5][h]
\draw[doubleline,gray] (-3.2,0) to [out=180, in=270] (-4.2,1.6) to [out=90,in =180] (-3.2,3.2) to (-.2,.2) to [out=315, in=180] (1,0);
\begin{scope}[xscale=-1,yscale=-1]
\draw[doubleline,gray] (-3.2,0) to [out=180, in=270] (-4.2,1.6) to [out=90,in =180] (-3.2,3.2) to (-.2,.2) to [out=315, in=180] (1,0);
\end{scope}
  \begin{scope}
\draw[doubleline] (0,2).. controls (2,2) .. (2,0.1);
\draw[doubleline] (2.1,0) -- (3,0);
\draw[doubleline] (0,-2).. controls (2,-2) ..  (2,-.1);
\draw[doubleline] (1.2,0) -- (1.9,0);
\end{scope}
\begin{scope}[xscale=-1,xshift=0cm]
\draw[doubleline] (0,2).. controls (2,2) .. (2,0.1);
\draw[doubleline] (2.1,0) -- (3,0);
\draw[doubleline] (0,-2).. controls (2,-2) ..  (2,-.1);
\draw[doubleline] (1.2,0) -- (1.9,0);
\end{scope}
\end{tikzpicture}
\end{center}

{\bf Diagram 9} with weight $N^2$ and multiplicity $2$ (a crossterm diagram):

\begin{center}
\begin{tikzpicture}[scale=0.4][h]
\draw[doubleline,gray] (-1,0) to [out=0,in=135] (0.9,1.1);
\draw[doubleline,gray] (-3.1,0) to [out=180,in=180] (-1,-3) to [out=0,in=215] (0.9,-1.1);
  \begin{scope}
\draw[doubleline] (0,2).. controls (4,2) .. (2,0.1);
\draw[doubleline] (1,1) -- (1.9,0.1);
\draw[doubleline] (0,-2).. controls (4,-2) ..  (2,-0.1);
\draw[doubleline] (1,-1) -- (1.9,-.1);
\end{scope}
\begin{scope}[xscale=-1,xshift=0cm]
\draw[doubleline] (0,2).. controls (2,2) .. (2,0.1);
\draw[doubleline] (2.1,0) -- (3,0);
\draw[doubleline] (0,-2).. controls (2,-2) ..  (2,-.1);
\draw[doubleline] (1.2,0) -- (1.9,0);
\end{scope}
\end{tikzpicture}
\end{center}

{\bf Diagram 10} with weight $N^2$  and multiplicity $2$ (a crossterm diagram). It equals Diagram 9 mirrored across the horizontal axis:

\begin{center}
\begin{tikzpicture}[scale=0.4][h]
\begin{scope}[yscale=-1]
\draw[doubleline,gray] (-1,0) to [out=0,in=135] (0.9,1.1);
\draw[doubleline,gray] (-3.1,0) to [out=180,in=180] (-1,-3) to [out=0,in=215] (0.9,-1.1);
  \begin{scope}
\draw[doubleline] (0,2).. controls (4,2) .. (2,0.1);
\draw[doubleline] (1,1) -- (1.9,0.1);
\draw[doubleline] (0,-2).. controls (4,-2) ..  (2,-0.1);
\draw[doubleline] (1,-1) -- (1.9,-.1);
\end{scope}
\begin{scope}[xscale=-1,xshift=0cm]
\draw[doubleline] (0,2).. controls (2,2) .. (2,0.1);
\draw[doubleline] (2.1,0) -- (3,0);
\draw[doubleline] (0,-2).. controls (2,-2) ..  (2,-.1);
\draw[doubleline] (1.2,0) -- (1.9,0);
\end{scope}
\end{scope}
\end{tikzpicture}
\end{center}

{\bf Diagram 11}  with weight $N^4$ and multiplicity $1$:

\begin{center}
\begin{tikzpicture}[scale=0.5][h]
\draw[doubleline,gray] (-0.9,1.1) to [out=35,in=145] (0.9,1.1);
\begin{scope}[yscale=-1]
\draw[doubleline,gray] (-0.9,1.1) to [out=35,in=145] (0.9,1.1);
\end{scope}
  \begin{scope}
\draw[doubleline] (0,2).. controls (4,2) .. (2,0.1);
\draw[doubleline] (1,1) -- (1.9,0.1);
\draw[doubleline] (0,-2).. controls (4,-2) ..  (2,-0.1);
\draw[doubleline] (1,-1) -- (1.9,-.1);
\end{scope}
\begin{scope}[xscale=-1,xshift=0cm]
\draw[doubleline] (0,2).. controls (4,2) .. (2,0.1);
\draw[doubleline] (1,1) -- (1.9,0.1);
\draw[doubleline] (0,-2).. controls (4,-2) ..  (2,-0.1);
\draw[doubleline] (1,-1) -- (1.9,-.1);
\end{scope}
\end{tikzpicture}
\end{center}

{\bf Diagram 12} with weight $N^2$ and multiplicity $1$:

\begin{center}
\begin{tikzpicture}[scale=0.5][h]
\draw[doubleline,gray] (-0.9,1.1) to [out=35,in=205] (0.9,-1.1);
\begin{scope}[yscale=-1]
\draw[doubleline,gray] (-0.9,1.1) to [out=35,in=205] (0.9,-1.1);
\end{scope}
  \begin{scope}
\draw[doubleline] (0,2).. controls (4,2) .. (2,0.1);
\draw[doubleline] (1,1) -- (1.9,0.1);
\draw[doubleline] (0,-2).. controls (4,-2) ..  (2,-0.1);
\draw[doubleline] (1,-1) -- (1.9,-.1);
\end{scope}
\begin{scope}[xscale=-1,xshift=0cm]
\draw[doubleline] (0,2).. controls (4,2) .. (2,0.1);
\draw[doubleline] (1,1) -- (1.9,0.1);
\draw[doubleline] (0,-2).. controls (4,-2) ..  (2,-0.1);
\draw[doubleline] (1,-1) -- (1.9,-.1);
\end{scope}
\end{tikzpicture}
\end{center}

In total, we thus have three diagrams with weights $N^4$, each with
multiplicity $1$. Moreover, we have nine diagrams with weights $N^2$,
three of which have multiplicity $1$, and six have multiplicity $2$.
This gives us a total relative weight of  
\begin{align}
\text{weight}= 3 N^4+ 15 N^2 \,.
\end{align}
The transition probability therefore equals
\begin{align}
\frac{1}{N^2} \Tr \, |{\cal T}_{12\rar 34}|^2 = w^2(3N^2+15) \,.
\end{align}
By demanding that this expression reproduces the result for $N=1$ (the theory of a single real scalar field), we find
$w^2= g^4 / 18$. 
The total transition probability is therefore
\begin{align}
\frac{1}{N^2} \Tr \, |{\cal T}_{12\rar 34}|^2 = \frac{g^4}{6}(N^2+5) \,,
\end{align}
which we used in the kinetic theory prediction, i.e. in Eq. (10), to give us the leading Lyapunov exponent $\lambda_L$.

\end{document}